\documentclass[journal=mamobx,manuscript=communication]{achemso}

\usepackage[version=3]{mhchem} 
\usepackage{subfigure}
\usepackage{color}

\author{Ting Ge}
\affiliation[Johns Hopkins University]{Department of Physics and Astronomy, Johns Hopkins University, Baltimore, MD 21218 USA}
\alsoaffiliation[University of North Carolina]{Department of Chemistry, University of North Carolina, Chapel Hill, NC 27599-3290, USA}
\author{Gary S. Grest}
\affiliation[Sandia National Laboratories]{Sandia National Laboratories, Albuquerque, NM 87185 USA}
\author{Mark O. Robbins}
\email{mr@jhu.edu}
\affiliation[Johns Hopkins University]{Department of Physics and Astronomy, Johns Hopkins University, Baltimore, MD 21218 USA}
\title[\texttt{achemso} Weld Crazing]
{Tensile Fracture of Welded Polymer Interfaces: Miscibility, entanglements and crazing}

\begin{document}


\begin{abstract}
Large-scale molecular simulations are performed to investigate tensile failure of polymer interfaces as a function of welding time $t$.
Changes in the tensile stress, mode of failure and interfacial fracture energy $G_I$ are correlated to changes in the interfacial entanglements as determined from Primitive Path Analysis.
Bulk polymers fail through craze formation, followed by craze breakdown through chain scission.
At small $t$ welded interfaces are not strong enough to support craze formation
and fail at small strains through chain pullout at the interface.
Once chains have formed an average of about one entanglement across the interface, a stable craze is formed throughout the sample.
The failure stress of the craze rises with welding time and the mode of craze breakdown changes from chain pullout to chain scission as the interface approaches bulk strength.
The interfacial fracture energy $G_I$ is calculated by coupling the simulation results to a continuum fracture mechanics model.
As in experiment, $G_I$ increases as $t^{1/2}$ before saturating at the average bulk fracture energy $G_b$.
As in previous simulations of shear strength,
saturation coincides with the recovery of the bulk entanglement density.
Before saturation, $G_I$ is proportional to the areal density of interfacial entanglements.
Immiscibiltiy limits interdiffusion and thus suppresses entanglements at the interface.
Even small degrees of immisciblity reduce interfacial entanglements
enough that failure occurs by chain pullout and $G_I << G_b$.

\end{abstract}

\begin{tocentry}
\includegraphics{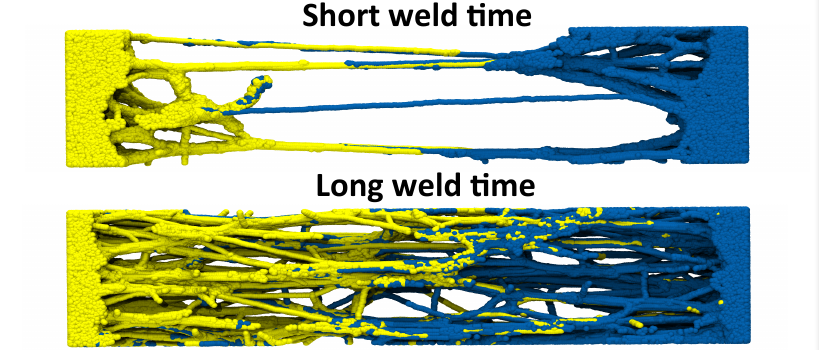}
\\
\\
\\\textbf{***For Table of Contents Use Only***}
\\\textbf{Title:} Tensile Fracture of Welded Polymer Interfaces:
\\\phantom{\textbf{Title:}} Miscibility, entanglements and crazing
\\\textbf{Authors:} Ting Ge, Gary S. Grest and Mark O. Robbins
\end{tocentry}

\section{1 Introduction}
\label{sec:intro}

The strength of polymer-polymer interfaces has great importance
in the application of polymers as adhesives, coating materials
and structural components \cite{wool95, haward97,jones99}.
Ideally the interface should be as strong as the surrounding bulk material
so that it does not initiate failure.
One common means of strengthening polymer interfaces is thermal welding \cite{wool95}.
The interface is heated above the glass transition temperature
so that polymers can diffuse across the interface.
While polymers must diffuse by a distance on the order of their radius of gyration for all memory of the interface to be erased, experiments \cite{wool95,schnell99,mcgraw13} and simulations \cite{ge13, ge14} have shown that bulk strength is recovered at much earlier times.

There has been great interest in understanding the evolution of interfacial
strength with interdiffusion time and its connection to interfacial structure \cite{schnell98,schnell99,brown01,jud81,prager81,prager83,wool81,kim83,mikos89,benkoski01,
cole03, mcgraw13,ge13,ge13b}.
Experiments \cite{schnell98,schnell99,brown01} have found a correlation between the mechanical strength 
and the interfacial width.
For homopolymers, the
growth of interfacial width with time has been explained using
reptation theory \cite{jud81,prager81,prager83,wool81,kim83,mikos89,benkoski01}.
For immiscible polymers, the
free energy cost of mixing limits interdiffusion,
causing both the interfacial width and strength to saturate \cite{helfand71, helfand72,cole03}.
Theories have interpreted the growth of strength with time in terms of entanglements \cite{jud81,prager81,prager83,wool81,kim83,mikos89,benkoski01,cole03}, but experiments can not directly image entanglements to test these predictions.

In recent work we have used simulations to follow the evolution of the
shear strength of interfaces between miscible \cite{ge13} and immiscible polymers \cite{ge13b}.
The maximum shear stress before failure $\sigma_{max}$ was evaluated using a
geometry that
mimics a lap-joint shear experiment \cite{kline88,parsons98,akabori06,boiko12}.
As in experiments \cite{kline88}, we found that $\sigma_{max}$ rose with welding time $t$
as $t^{1/4}$ and saturated long before polymers had diffused by their radius
of gyration.
In contrast to experiments,
simulations allowed direct observations of the evolution of topological constraints (TCs) associated with entanglements.
The shear strength rose linearly with the areal density of interfacial entanglements between polymers that started on opposite sides of the interface.
Homopolymers achieved bulk strength when they had interdiffused enough
to produce roughly two TCs per chain.
This was sufficient to prevent chain pullout at the interface and led to failure through chain scission, as in the bulk.
Even a small degree of immiscibility prevented interfacial entanglements,
resulting in weak interfaces that failed through chain pullout.

In many applications polymer interfaces are under tensile loading and the
ultimate mechanical strength of the whole system containing them
is determined by their interfacial fracture energy $G_I$ \cite{wool95,haward97,jones99}.
The fracture energy is the amount of external work per unit area required to propagate a crack across the interface
 \cite{haward97}. It represents an integral of the stress times displacement and rises both with the peak tensile stress and the total deformation of the system.
A lower bound for $G_I$ is the free energy change $2\gamma$, where $\gamma$ is the interfacial free energy.
Common methods of measuring $G_I$ include the
double cantilever beam experiment and T-peel test \cite{wool95,schnell98,schnell99,brown01,cole03}.
Experiments on homopolymer interfaces \cite{jud81} show that $G_I$ grows as $t^{1/2}$, which is proportional to the interfacial width, before saturating at the bulk fracture energy $G_b$.

The bulk fracture energy of glassy polymers is typically thousands of times
larger than $2\gamma$. This increase is not due to a large increase in the peak tensile stress, but rather because a large volume around the crack tip is deformed into a network of fibrils and voids called a craze \cite{haward97,donald82a,donald82b,kramer83,kramer90}. The density of craze is lower
than that of the bulk glass by the extension ratio $\Lambda$.
In most cases $\Lambda$ correlates to the entanglement molecular weight of
the polymer and is insensitive to temperature, strain rate and other conditions \cite{donald82a, donald82b,kramer83,kramer90,haward97}.
Formation of the craze occurs by drawing fibrils out of the
bulk polymer at a constant plateau stress $S$.
These oriented fibrils are then strong enough to resist further deformation.
Brown \cite{brown91} developed a fracture mechanics model for $G_I$ in terms of $\Lambda$, $S$ and the ultimate failure strength of fully developed crazes $S_{max}$.
Rottler et al.\cite{rottler02} later used molecular dynamics (MD) simulations to determine the inputs to Brown's model.
Their results capture many aspects of craze formation,
including the large fracture energies of bulk polymers.

In this paper we use simulations to study the tensile strength of the interfaces between miscible and immiscible homopolymers whose shear strength was studied in Refs. \citenum{ge13} and \citenum{ge13b}.
For miscible polymers at small times, and for the immiscible
interfaces studied here,
the interface separates easily and chains are pulled back to their initial side of the interface with a small interfacial fracture energy.
The fracture energy begins to rise rapidly when the interface is strong enough to allow formation of a stable craze over a large volume. This coincides with formation of the first TCs of polymers with polymers on the opposing side and occurs long before there is a significant increase in the peak shear or tensile stress.
Brown's theory is used to determine $G_I$, which grows as $t^{1/2}$ for miscible polymers.
Final saturation of strength occurs when interdiffusion has restored the entanglement density near the interface to its bulk value.
This saturation occurs at the same time as the saturation of shear strength \cite{ge13}.
While $\sigma_{max}$ only increases by a factor of 2 or 3 during welding,
$G_I$ increases by more than two orders of magnitude because of the large increase in craze volume.
As in studies of shear strength\cite{ge13b}, interfaces between immiscible
polymers are weaker because their
finite equilibrium width suppresses entanglements.
The equilibrium fracture energy is comparable to that of homopolymers at welding
times that produce the same density of interfacial entanglements.


The following section describes the simulation geometry and method for determining fracture energy.
In Section 3, results are presented first for miscible polymers and then for the immiscible case.
The final section presents a summary and conclusions.

\section{2 Simulation Model and Methodology}
\label{sec:method}

The potential models and welded states were the same as in Refs.\citenum{ge13}, \citenum{ge14} and \citenum{ge13b} where further details of their preparation can be found.
All of the simulations employed the canonical bead-spring model \cite{kremer90} that captures the properties of linear homopolymers.
The van der Waals interactions between spherical monomers of mass $m$ are
modeled using the standard Lennard-Jones (LJ) potential with interaction energy $u_0$,
monomer diameter $a$ and characteristic time $\tau=a\sqrt{m/u_0}$.
The potential was truncated and shifted to zero at monomer separation $r_c=1.5a$ or $2.5a$.
To model immiscible films, the interaction strength $u_0$ between unlike monomers was reduced to $\tilde{\epsilon}_{12}u_0 <u_0$.
Four systems with $\tilde{\epsilon}_{12}=1.0$, $0.99$, $0.98$ and $0.95$ were simulated.

Chains of length $N=500$ beads were made by connecting nearest-neighbors using an additional bonding potential.
Melt simulations were performed with the usual unbreakable finitely extensible nonlinear elastic (FENE) bonding potential\cite{kremer90}.
A simple quartic potential with the same equilibrium spacing was used in the mechanical tests since chain scission plays an essential role in failure.
As in past simulations \cite{ge13,ge13b,ge14,rottler02,stevens01},
the breaking force for the quartic potential, $240 u_0/a$, is 100 times higher than that for the intermolecular LJ potential.
This ratio is consistent with  experimental estimates \cite{odell86,creton92,ge14}.
Previous work has shown that the entanglement length for this model is $N_e = 85 \pm 7$ and that the mechanical response for $N=500$ is characteristic of highly entangled (large $N$) polymers \cite{rottler02, rottler02b, rottler03, hoy07, hoy08,ge14}.

There is no precise mapping of our coarse-grained model to any specific polymer, but approximate mappings give $u_0/a^3 \sim 50$MPa and $u_0/a^2 \sim 25$mJ/m$^2$
for the units of stress and fracture energy, respectively \cite{ge14}.
The interfacial energy of the polymer-vacuum interface
is
$\gamma \sim 1 u_0/a^2 \sim 25 $mJ/m$^2$
with a weak dependence on temperature \cite{gersappe99,baljon01}.
The fracture energy must exceed the free energy associated with the two
interfaces created, $G_I > 2\gamma$.

Fluid films of each polymer species were equilibrated separately at temperature $T=1.0u_0/k_B$ with $r_c=2.5a$.
Each film contains $M=4800$ chains or in total $2.4$ million beads.
Periodic boundary conditions were applied along the $x$- and $y$-directions with dimensions $L_x=700a$ and $L_y=40a$, while featureless walls separated by $L_z=100a$ confined films in the non-periodic $z$- direction.
Equilibrated films were placed in contact and allowed to interdiffuse for a time $t$ at $T=1.0u_0/k_B$.
The system was then quenched rapidly to $T=0.2 u_0/k_B$, which is below the glass temperature $T_g \approx 0.35 u_0/k_B$ \cite{rottler03c}.
Glass simulations are done with $r_c=1.5a$ to reduce the difference between melt and glass densities, and thus changes in chain conformation\cite{foot1}.
This also facilitates comparison with previous mechanical studies \cite{rottler02,rottler02b,rottler03,hoy07, hoy08}.
The cutoff radius was reduced to $r_c=1.5a$ prior to quenching.
The system was then quenched 
at constant volume with a rate $\dot T = -10^{-3} u_0/(k_B\tau)$ to $T=0.5u_0/k_B$ where the pressure $P=0$. Subsequently, the temperature was further quenched to $T=0.2 u_0/k_B$ at $\dot T = - 2 \times 10^{-4} u_0/(k_B\tau)$ and $P=0$. 

Entanglements in all quenched states were identified using the Primitive Path Analysis (PPA) algorithm \cite{everaers04}, which has provided unique information about the average spacing between entanglements both in a polymer melt at equilibrium \cite{everaers04} and during the recovery of the entanglement network from a non-equilibrium situation \cite{vettorel10}.
During PPA, chain ends are frozen and tensile forces are introduced to minimize the contour length without allowing chain crossing. 
To limit excluded volume effects, we use a modification of the original PPA algorithm where the diameter of repulsive interactions between chains is then reduced by a factor of four and the contour is minimized again \cite{hoy07b,ge13,ge13b}.
The resulting configuration is a network of primitive paths for each chain. 

As in previous studies \cite{hoy07b,tzoumanekas06,everaers12,ge13,ge13b},
we identified the interchain contacts between primitive paths as topological constraints (TCs) associated with entanglements.
The spacing between TCs is typically 2 to 3 times smaller \cite{tzoumanekas06,everaers12} than the entanglement length, which is defined as the Kuhn length of the primitive path. The reason is that several contacts with primitive paths of other chains are needed to randomize the direction. Recent studies of entangled polymer melts \cite{anogiannakis12} and craze formation \cite{ge13thesis} also show that unlike TCs, entanglements are not associated with specific pairs of chains. However, all of the past work has found that the densities of TCs and entanglements are proportional to each other. Refs. \citenum{ge13} and \citenum{ge13b} describe the evolution of the ratio of the local density of TCs to the bulk density. This is equivalent to the evolution of the entanglement density and we will refer to these results below. We also present new measures of the evolution of TCs that are related to the onset of interfacial craze formation.

\begin{figure}[ptb]
\begin{center}
\includegraphics[width=0.50\textwidth]{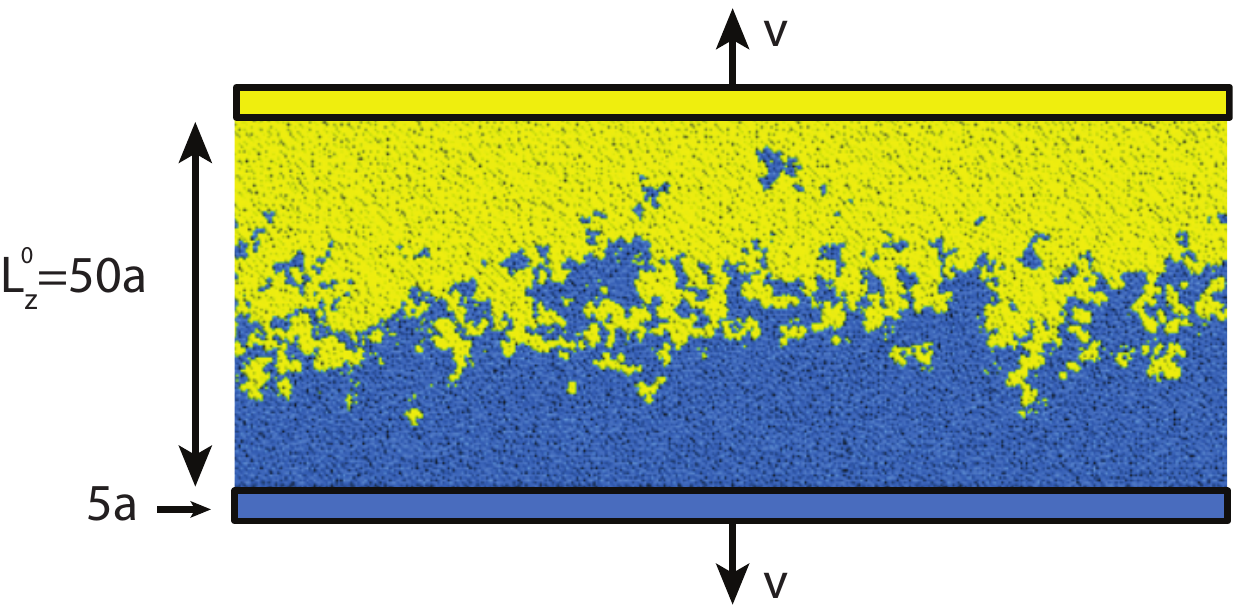}
\end{center}
\caption{Schematic of tensile test. Monomers in top and bottom layers of width $5a$ are displaced in opposite directions at speed $v$. They impose a strain on a region centered on the initial interface whose initial width is $L_z^0 = 50a$. Polymer beads are colored yellow and blue if their initial positions before interdiffusion are at $z>0$ and $z<0$, respectively.  
}
\label{fig:schematic}
\end{figure}

To perform tensile tests that isolate the effect of the interface \cite{ge14} we apply strain to a region of height $L_z^0=50 \sigma$ centered on the initial interface.
Layers of width $5a$ above and below this region are held rigid
and displaced at constant velocity $v=0.01a\tau^{-1}$ in opposite directions along the $z-$axis. Our simulation protocol is schematically illustrated in Figure \ref{fig:schematic}.
The deformation is characterized by the stretch factor along the $z-$axis $\lambda\equiv L_z/{L_z}^0$, where $L_z$ is the growing separation between the top and bottom rigid layers.
The periodic boundary conditions along $x$ and $y$ directions are not changed.
The temperature is maintained at $T=0.2u_0/k_B$ by applying a Langevin thermostat with damping rate $\Gamma=1\tau^{-1}$ to the peculiar velocities in $x$ and $y$ directions. 

Tests with different $L_z^0$, $L_x$, $L_y$, $v$ and $\Gamma$ gave similar results.
As discussed in Ref. \citenum{ge14}, the rigid walls screen the stress on regions separated by more than $L_z$ along the x-y plane.
Thus the data for different regions along $L_x=700a$ are independent and the total stress from a single run is effectively an ensemble average over roughly 14 regions.
Previous simulations \cite{rottler03,baljon01} showed that simulations of craze formation should be done at constant widening velocity rather than constant true strain rate because deformation is localized to an active zone of fixed
width at the edge of the craze.
Rottler and Robbins \cite{rottler03} examined the velocity dependence of the plateau stress $S$ during craze formation and found a weak, logarithmic variation
for $v < 0.03 a\tau^{-1}$.

To obtain the average tensile response of bulk samples with chain length $N=500$, we let one polymer film evolve at constant temperature $T=1.0 u_0/k_B$ and constant volume.
We then selected six states at times separated by equal intervals, $\Delta t = 50 k\tau$, for mechanical testing. Simulation protocols for quenching and tensile tests on these bulk samples were identical to those for polymer interfaces. 

\section{3 Results}
\label{sec:result}

\subsection{3.1 Fully Miscible Interface}
\label{sec:miscible}
\subsubsection{3.1.1 Evolution of Stress and Failure Mechanism}
\label{sec:stress}

\begin{figure}[htp]
\includegraphics[width=0.85\textwidth]{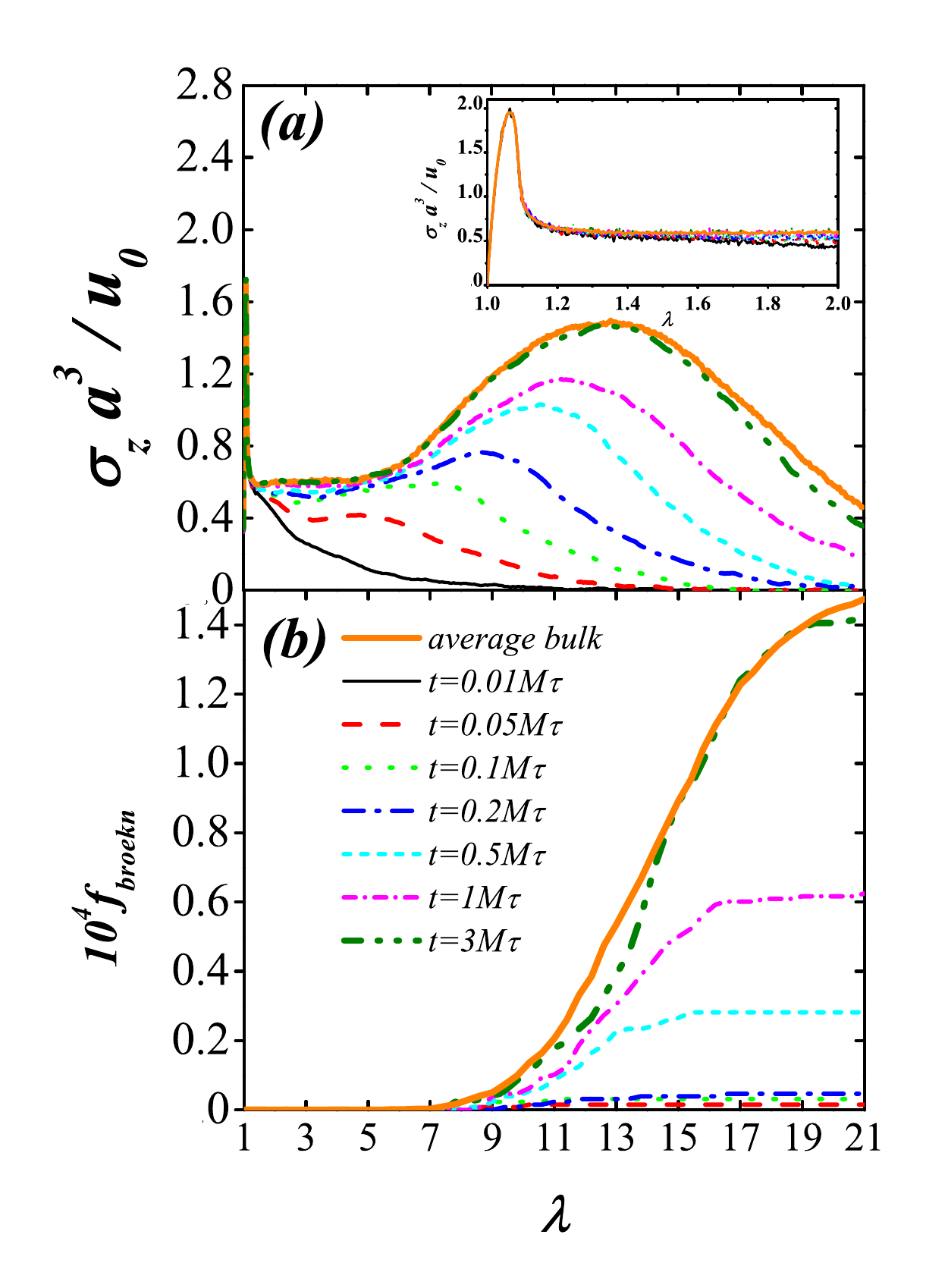}
\caption{(a) Tensile stress $\sigma_z$ and (b) fraction of broken bonds $f_{broken}$ as a function of the stretch factor $\lambda$ for the interface between fully miscible polymers at the indicated interdiffusion times $t$, and for the bulk. The inset of (a) shows the tensile stress $\sigma_z$ at stretch factor $1<\lambda<2$.}
\label{fig:stress}
\end{figure}

\begin{figure}[tbp]
\begin{center}
\includegraphics[width=0.50\textwidth]{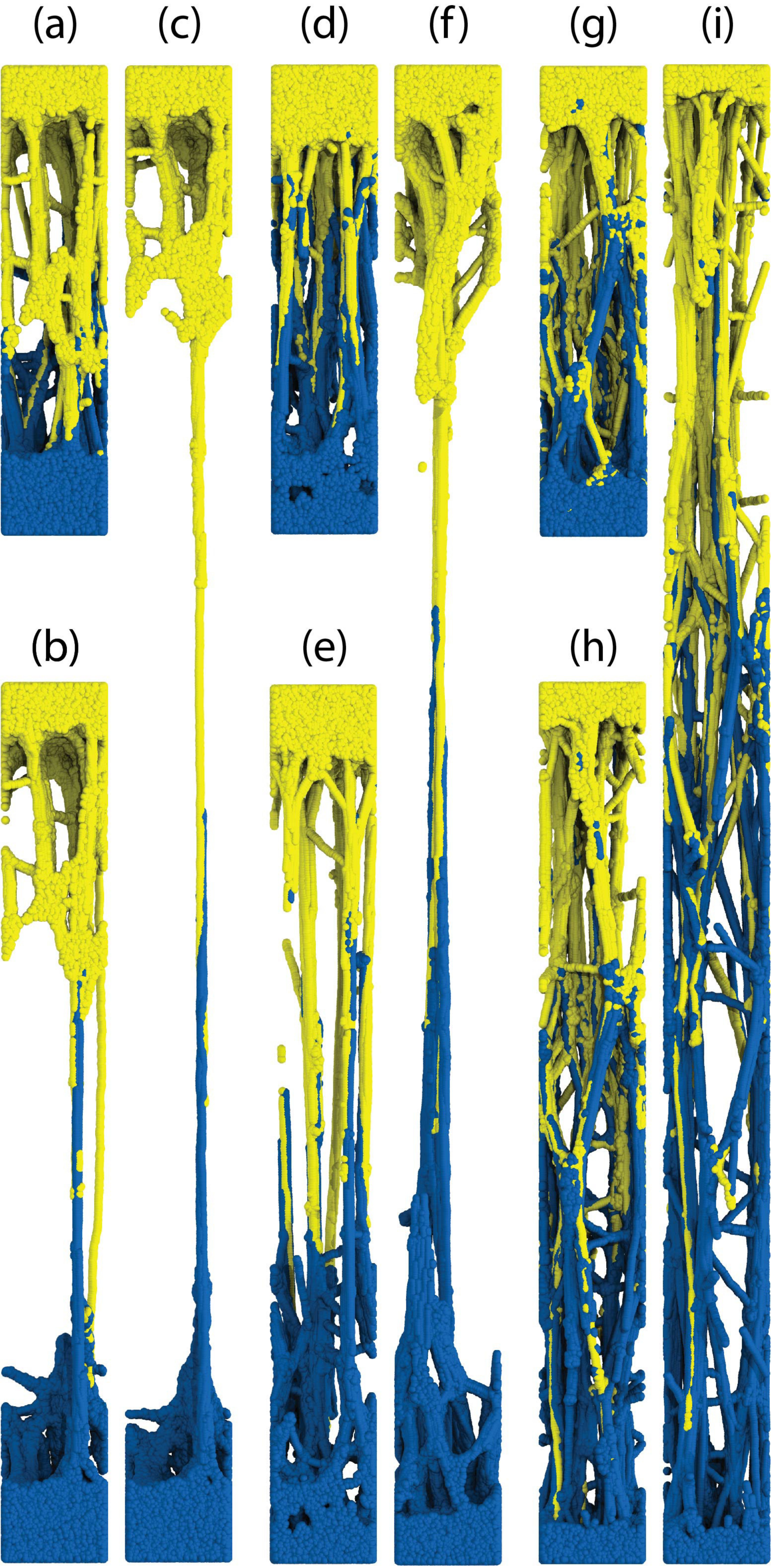}
\end{center}
\caption{Snapshots of the interface between fully miscible polymers during
tensile tests at $T=0.2u_0/k_B$ for $t=0.01M\tau$ ((a)-(c)), $0.05M\tau$ ((d)-(f)) and $5M\tau$ ((g)-(i)). For each $t$, results are shown for $\lambda=3.4$, $6.6$ and $11.4$. Beads are colored based on their positions before the interdiffusion: $z>0$ (yellow) and $z<0$ (blue). 
For clarity, only a portion of the sample $40a$ deep in the x-direction (into the page) is shown. Chains that appear to end are just leaving the front, back or sides of the pictured region.
}
\label{fig:snapshot}
\end{figure}


Figure \ref{fig:stress}(a) shows the tensile stress $\sigma_z$ as a function of the stretch factor $\lambda$ for interfaces between fully miscible chains at the indicated interdiffusion times $t$. A million $\tau$ is abbreviated as 1M$\tau$ and 
the average result for bulk films is shown for comparison.

As shown in the inset of Fig.\ref{fig:stress}(a), for $1 < \lambda < 1.1$ the stress curves show a linear elastic region followed by a peak and sharp drop, 
indicating the start of cavitation.
The similarity between the curves for different $t$ confirms that 
cavitation is controlled by the local structure and stress state \cite{haward97,rottler03}, which is nearly identical for all $t$ and the bulk.
A consequence is that cavitation and subsequent craze nucleation can occur at places away from the interface.
However, because of the thin film geometry in our simulations, the results mainly
reflect the effects of the interface on the fracture process.
In experiment, examining purely interfacial fracture is challenging.
Special techniques are usually required to confine the fracture process as near to the interface as possible \cite{wool95, schnell98,schnell99}.

Changes in the mechanical response become evident after cavitation
($\lambda > 1.1$).
For $t<0.1M\tau$, $\sigma_z$ drops to zero without a clear stress plateau, indicating that the system fails without forming a stable craze.
As shown in Fig.~\ref{fig:stress}(b), for these states there are almost no broken bonds during the whole process. Instead, the interface fails through chain pullout. Figures~\ref{fig:snapshot}(a)-(f) illustrate the evolution of systems at $t=0.01M\tau$ and $0.05M\tau$ upon stretching to $\lambda=3.4$, $6.6$ and $11.4$. These snapshots show that a few fibrils can form across the interface, but quickly fail through pullout.
In the final state (not shown) all monomers are on the side they started from before interdiffusion.

For $t\geq0.1M\tau$, there is a pronounced plateau in $\sigma_z$ at the average bulk plateau stress $S\sim0.6u_0/a^3$.
This indicates that a stable craze has formed and grows steadily through the interfacial region \cite{kramer83,kramer90,rottler03}.
Figure~\ref{fig:snapshot}(g) illustrates the structure of the craze, an intricate network of fibrils and voids.
The density of the craze is lower than the initial density by the same extension
ratio $\Lambda \approx 8$ that is found for bulk samples \cite{rottler03,ge14,ge13thesis}.
The stress plateau ends when all the material between the two rigid layers, except regions within a distance on the order of the tube diameter from the rigid layers, has been transformed into craze.
Because of these uncrazed regions,
the plateau ends at a value of $\lambda \sim 6$ that is smaller than $\Lambda$.
Figure~\ref{fig:stress}(b) shows that there is no bond breaking during craze
formation.

Upon further stretching of the fully developed craze, $\sigma_z$ rises above $S$ and reaches a maximum stress $S_{max}$ before the ultimate breakdown of the craze.
For $t=0.1M\tau$, $S_{max}$ is close to $S$, but $S_{max}$ rises rapidly as $t$ increases.
The failure mechanism of craze fibrils also changes from chain pullout to chain scission.
For $t=0.1M\tau$ and $0.2M\tau$, only a few bonds break.
For $t>0.2M\tau$, significant bond breaking is observed.
The peak rate of bond breaking and the final fraction of broken bonds increase with $t$.
In each case, the rate of bond breaking begins to drop near the point where $\sigma_z$ reaches $S_{max}$.
Deformation of the craze in this range of times is illustrated by Figures~\ref{fig:snapshot}(h) and (i) for $t=5M\tau$.
Snapshot (h) shows the structure at $\lambda=6.6$, where the entire system is stretched into a uniform craze without
significant bond breaking.
Snapshot (i) shows the same interfacial craze at $\lambda=11.4$, where the ultimate breakdown has started.
Note that chains from the two opposite sides remain mixed rather than being pulled out to their initial sides, as for small $t$.
Around $t=3M\tau$, the rate of bond breaking and the entire stress curve become statistically indistinguishable from the bulk result.

\subsubsection{3.1.2 Correlating Failure Mechanisms to Entanglements}
\label{sec:entanglements}

In the previous section we identified several characteristic times where the mechanical response changes.
There is an onset of stable craze formation at $t_c \sim 0.1$M$\tau$, a transition from chain pullout to chain scission at $t_s \sim 0.5$M$\tau$ and a saturation at the bulk response for $t \geq t_{b} \sim 3$M$\tau$.
As in our previous studies of shear strength \cite{ge13, ge14}, these 
transitions correlate with the evolution of entanglements near the interface as measured by identifying TCs between the primitive paths of different chains.

Previous simulations \cite{baljon01,rottler03} showed that in mono-dispersed bulk samples
stable crazes began to form when the chain length was increased from 128 to 256.
To sharpen this criterion, additional simulations of bulk samples were performed for a set of chain lengths in the range $25\leq N \leq 500$.
We found that homopolymers with chains of length $N\leq175$ did not form a stable craze, while those with $N\geq200$ did \cite{ge13thesis}.
Since $N_e\sim 85$, craze growth requires roughly two entanglements per chain.
At small welding times, chains have not diffused far enough to form these entanglements
and the interface responds like a bulk system with $N<175$.


 
Results for the density of entanglements near the interface as a function of welding time are presented in Ref. \citenum{ge13}.
They show that at $t_c$ the interfacial entanglement density is still strongly suppressed from the bulk value for chains with $N=500$.
To identify a criterion for stable craze formation we examined a new measure,
the density of entangled interfacial strands.
These are defined as
chain segments that cross the mid-plane $z=0$ once and have one end forming a TC with a chain from each side.
Based on the bulk results described above, we expect that this is a minimal criterion
for the chain segment to contribute to stable craze formation. 

\begin{figure}[tbp]
\begin{center}
\includegraphics[width=0.75\textwidth]{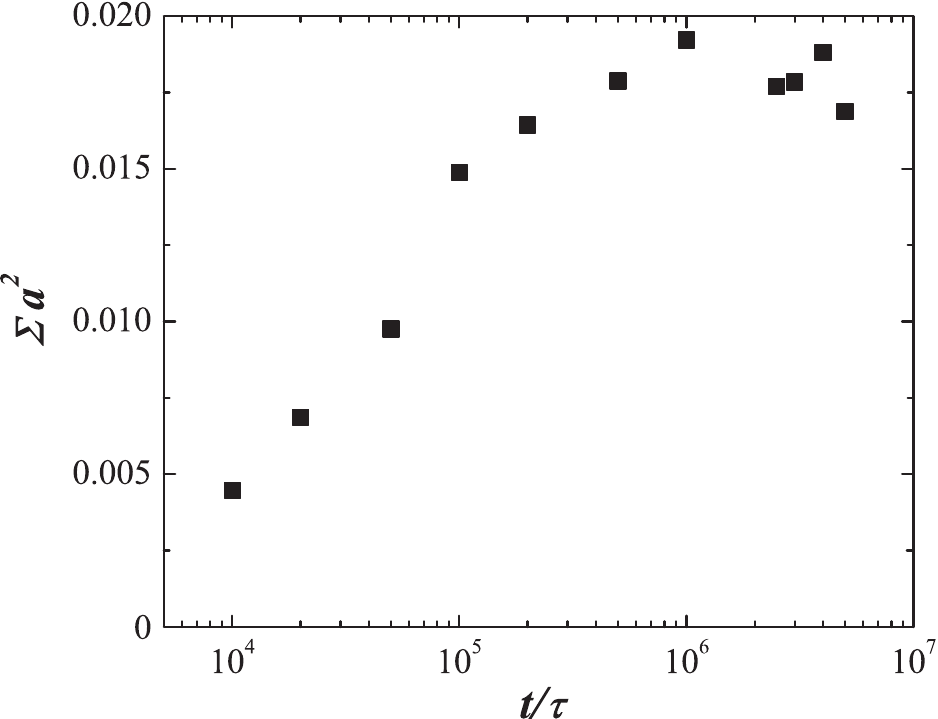}
\end{center}
\caption{Development of the areal density $\Sigma$ of interfacial entangled strands with interdiffusion time $t$ during thermal welding of fully miscible polymers.
}
\label{fig:strand}
\end{figure}

Figure~\ref{fig:strand} shows the time dependence of the areal density
of entangled interfacial strands, $\Sigma$.
At small times, $\Sigma$ rises as chains diffuse across the interface.
For $t > 0.2M\tau$, $\Sigma$ saturates.
There is no further increase in the number of chains that cross the interface
to form entanglements.
The total number of interfacial entanglements between chains from opposite sides continues to grow \cite{ge13}, but only because chains that cross the interface
diffuse farther and form more entanglements on the opposing side.
Note that $\Sigma$ saturates near $t_c$.
Thus, once chains have formed the first TC across the interface, chain pullout is suppressed and a craze is formed.

For $t > t_c$ the interface and bulk deform into a craze in the same way.
Mechanical stability requires that the tensile stresses are the same and we find the
final densities correspond to the same extension ratio $\Lambda$.
For bulk crazes, both experiment \cite{donald82a, donald82b, kramer83,kramer90} and simulation \cite{rottler03} results for $\Lambda$ are consistent with a model based on the idea that segments between
entanglements become taut during crazing.
We verified that entangled 
interfacial strands stretched by the same amount as their bulk counterparts.
However, as in previous work \cite{rottler03,rottler02b,ge13thesis}, the strands are not completely taut.
This reflects a factor of $\sqrt{3}$ from an average over orientations that was dropped from the original derivation of the expression for $\Lambda$ \cite{donald82a, donald82b,rottler02b}.

In mono-dispersed samples, previous simulations \cite{rottler03} have shown that
the mechanism of craze failure changes with increasing chain length.
For $2N_e < N<3N_e$, the craze forms and then fails through chain pullout, while for $N>3N_e$ the craze fails through chain scission. 
This criterion again highlights the role of entanglements in the breakdown of crazes.
Our previous analysis \cite{ge13} has shown that the density of entanglements keeps rising at the interface for $t > t_c$.
At $t=0.5M\tau$, we can already observe a considerable amount of bond breaking during craze failure (Fig.~\ref{fig:stress}(b)).
The bulk mechanical response (and accordingly the level of bond breaking) is recovered when the bulk entanglement density is recovered at the interface.
This occurs by $t=3M\tau$, which is the same time where the bulk shear strength is recovered \cite{ge13}.
Shear studies also found a rapid onset of bond breaking between $t=0.2M\tau$ and $0.5M\tau$. 

The time for diffusion to completely erase the initial interface should be of order the disentanglement time $\tau_d \sim 30$M$\tau$ for chains to diffuse by their end-end distance \cite{ge14}. 
Our finding that bulk behavior is recovered for $t << \tau_d$ is consistent
with several previous studies.
Tensile tests by Schnell et al. \cite{schnell99} also showed that strength recovers at times much smaller than $\tau_d$.
More recently, McGraw et al. \cite{mcgraw13} examined
crazes formed at the interface between two identical polymer films
as a function of welding time \cite{mcgraw13}.
They found that the extension ratio of the interfacial craze reaches that of the bulk craze at $t << \tau_d$.
In Ref. \citenum{ge13} we showed that 3M$\tau$ corresponds to the time for
the density of entanglements near the interface to return to the bulk
value.
Chains have only diffused far enough to create roughly two entanglements per chain with chains on the other side of the interface,
but this is enough to prevent chain pullout and achieve bulk strength.

\subsubsection{3.1.3 Macroscopic Fracture Energy}
\label{sec:fracture}

The interfacial tensile strength is usually characterized by the interfacial fracture energy $G_I$ which corresponds to the work needed to fracture a unit area \cite{wool95,schnell98,schnell99,brown01,cole03}.
For systems that do not form a stable craze ($t<0.2M\tau$),
deformation is localized to the interface.
$G_I$ can then be obtained by integrating the work under the stress curves in Figure~\ref{fig:stress}: $G_I = \int\sigma_z\,L_z^0 d\lambda $.
This method is not sufficient for $t\geq0.2M\tau$,
because the rigid boundary layers limit the growth of the craze and therefore prevent the simulation from capturing the whole plastic zone that would form ahead of an experimental crack tip.
To circumvent this limitation, we apply the same method as in a previous study \cite{rottler02}, where Brown's fracture model and molecular simulation were combined to estimate the fracture energy for polymer glasses.

Brown's fracture model \cite{brown91} links two length scales: the macroscopic scale where the fracture test is performed and the microscopic scale of fibrils and voids within the craze.
In his model, $G_I=Sd(1-1/\Lambda)$, where $d$ is the maximum width of the craze ahead of the crack tip.
Brown showed that $d=4\pi\kappa(S_{max}/S)^2 D_0$, where
$D_0$ is the spacing between fibrils and the dimensionless $\kappa$ can be expressed in terms of various elastic moduli of the fully developed craze.
Since $S_{max}$ is not directly accessible to experiments, Brown estimated $S_{max}$ from the stress needed for chain scission, though the model applies for any mode of craze failure. 

\begin{figure}[tbp]
\begin{center}
\includegraphics[width=0.75 \textwidth]{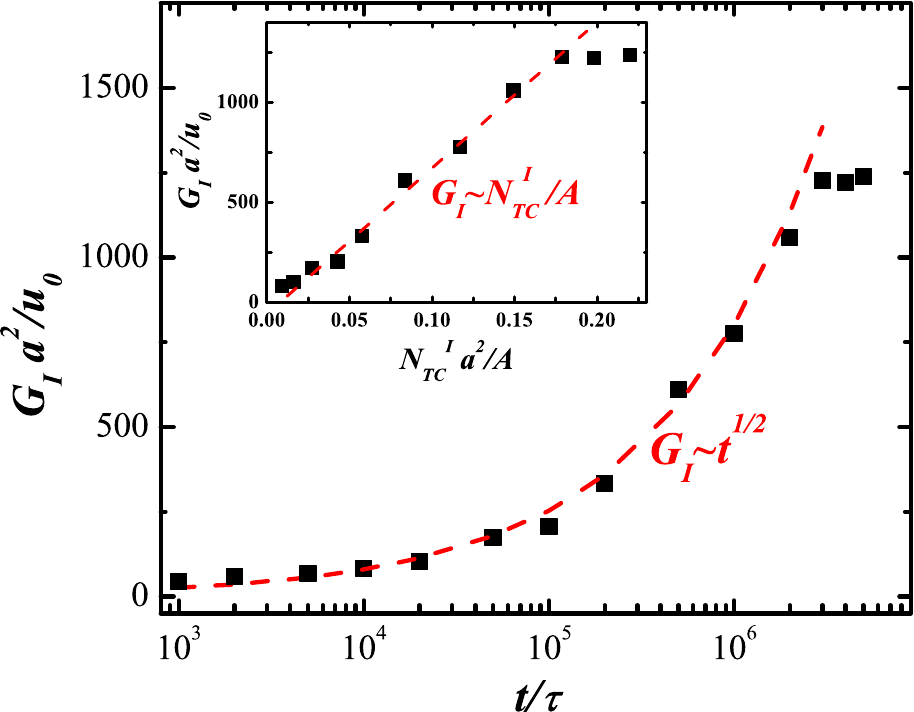}
\end{center}
\caption{Development of the interfacial fracture energy $G_I$ with 
welding time $t$.
The dashed line indicates the $t^{1/2}$ power law.
The inset plots $G_I$ against $N_{TC}^I/A$, the areal density of interfacial TCs.
The dashed line indicates the linear correlation between $G_I$ and $N_{TC}^I/A$ before $G_I$ saturates at $G_b$.
}
\label{fig:energy}
\end{figure}

All quantities needed for the calculation of $G_I$ using Brown's model can be measured in our simulation.
For $t>0.2M\tau$ the plateau stress and extension ratio are nearly
constant, $S=0.6u_0/a^3$ and $\Lambda=8$.
The ratio $S_{max}/S$ is extracted from stress curves like those shown in Fig.~\ref{fig:stress}.
Previous simulations \cite{rottler02} using the same polymer model determined that $\kappa=2.0-2.8$ and $D_0=10-14a$.
Here, $D_0\sim12a$ and $\kappa\sim2.5$ are used in our calculation of $G_I$. 

The increase of $G_I$ with $t$ is shown in Figure~\ref{fig:energy}.
At long times, $G_I$ approaches the bulk fracture energy $G_b$.
Both are about $1000$ times higher than the lower bound for $G_b$ that is given by twice the surface tension $\gamma \sim 0.6u_0/a^2$.
This ratio is consistent with experimental observations that crazing increases the fracture energy by factors of several thousand \cite{haward97}.
As in experiment \cite{jud81,wool95}, the increase of $G_I$ with $t$ follows a roughly $t^{1/2}$ power law before saturating at the bulk fracture energy $G_b$.
A reptation \cite{degennes71} argument has been employed to explain this power law based on formation of entanglements across the interface.
However, a recent simulation  \cite{pierce11} has shown that the dynamics across the interface between entangled melts is dominated by chain ends, and cannot be simply described as a reptation process.
Differentiating between these pictures is difficult given the limited scaling range in experiments.

Our simulation allows a direct test of the connection between $G_I$ and entanglements.
The inset of Fig.~\ref{fig:energy} shows a clear linear correlation between $G_I$ and $N_{TC}^I/A$, the areal density of interfacial TCs, before $G_I$ saturates at the bulk value.
The proportionality between $G_I$ and $N_{TC}^I/A$ can be related to the failure mechanism of chain scission, which is fostered by entanglements.
Note that this correlation is independent of the exact scaling of $N_{TC}^I/A$ or $G_I$ with $t$.

\subsection{3.2 Immiscible Interface}
\label{sec:immiscible}

\begin{figure}[ptb]
\begin{center}
\includegraphics[width=0.75\textwidth]{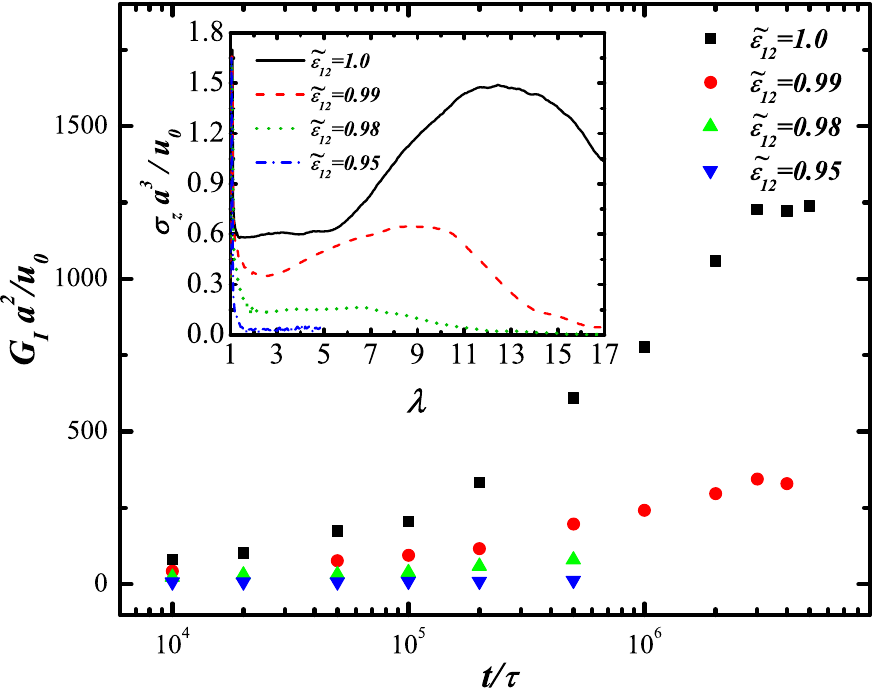}
\end{center}
\caption[Interfacial fracture energy vs. interdiffusion time and tensile stress-strain curve for different polymer interfaces.]{Development of $G_I$ with $t$ at the fully miscible ($\tilde{\epsilon}_{12}=1.0$) and immiscible ($\tilde{\epsilon}_{12}<1.0$) interfaces. The inset shows $\sigma_z$ vs. $\lambda$ for the fully miscible interface at $5M\tau$ and the immiscible interfaces at equilibrium.
}
\label{fig:energy2}
\end{figure}

For immiscible polymer interfaces, the free energy cost of mixing limits the degree of interdiffusion.
The interfacial width grows more slowly with time than for the miscible case and saturates at a finite value at large times \cite{ge13b}.
The growth in fracture energy with $t$ reflects these differences,
as shown in Fig.~\ref{fig:energy2}.
Immiscible results are always below those for the miscible case.
For $\tilde{\epsilon}_{12}=0.99$, $G_I$ saturates at $t \sim 3$M$\tau$.
For more immiscible cases, $G_I$ shows almost no increase and saturates by $0.5$M$\tau$.

\begin{figure}[ptb]
\begin{center}
\includegraphics[width=0.50\textwidth]{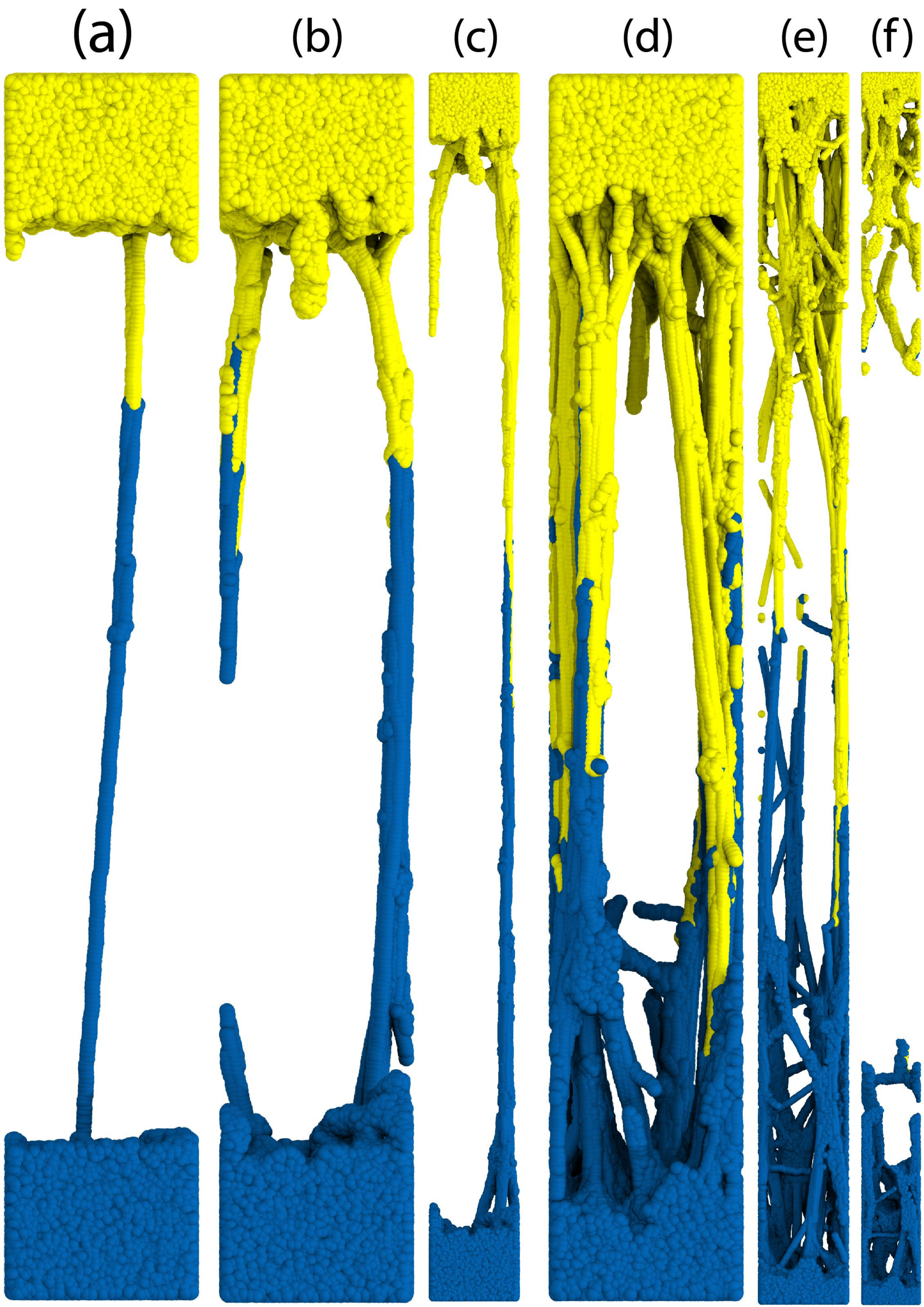}
\end{center}
\caption{Visualization of failure of equilibrated immiscible interfaces during tensile tests.
Snapshots correspond respectively to $\tilde{\epsilon}_{12}=0.95$ at $\lambda=5$ (a), $\tilde{\epsilon}_{12}=0.98$ at $\lambda=5$ (b) and $11$ (c), and $\tilde{\epsilon}_{12}=0.99$ at $\lambda=5$ (d), $11$ (e) and $17$ (f).
Beads are colored based on their positions before the interdiffusion: $z>0$ (yellow) and $z<0$ (blue).
For clarity, only a portion of the sample $40a$ deep in the x-direction (into the page) is shown. Chains that appear to end are just leaving the front, back or sides of the pictured region.
}
\label{fig:snapshot2}
\end{figure}

The inset of Fig.~\ref{fig:energy2} shows stress-strain curves for immiscible interfaces at their equilibrium state and for the miscible case at $5$M$\tau$.
The trends with increasing $\tilde{\epsilon}_{12}$ are similar to those with increasing time for miscible interfaces in Fig.~\ref{fig:stress}.
The interface fails rapidly after cavitation for
$\tilde{\epsilon}_{12}=0.95$ and $0.98$.
Figures~\ref{fig:snapshot2}(a-c) show that these systems fail through chain pullout
to the original side with no craze formation.
For $\tilde{\epsilon}_{12}=0.99$ the stress has a value close to the bulk plateau stress for a range of strains. 
Figures~\ref{fig:snapshot2}(d-f) show some fibrils, but there is no stable craze and failure is through chain pullout.

In Ref. \citenum{ge13b} we showed that a minimum interfacial width 
was needed for entanglements to form across the interface.
The equilibrium width for $\tilde{\epsilon}_{12}\leq 0.98$
is below this threshold, which is consistent with the rapid failure shown in Fig.~\ref{fig:energy2}.
For $\tilde{\epsilon}_{12}=0.99$ there are a few interfacial entanglements at $t>0.2M\tau$ and $G_I$ rises noticeably after this time.
However the equilibrium density of entanglements at long times is far below the bulk value needed for significant shear strength \cite{ge13b}.

For the miscible case we found that stable craze formation
was associated with saturation of the areal density of interfacial strands
at $\Sigma \sim 0.018 a^{-2}$.
For $\tilde{\epsilon}_{12}=0.99$, $0.98$ and $0.95$,
$\Sigma$ at the equilibrium interface is $\sim 72\%$, $34\%$ and $12\%$ of the saturation value needed for stable craze formation.
In all cases the equilibrium $G_I$ is comparable to that for miscible interfaces at the time when they have reached a comparable $\Sigma$.

\section{4 Summary and Conclusions}
\label{sec:sum}

We have presented detailed studies of the evolution with welding time of interfacial entanglements and the tensile fracture energy of welded polymer interfaces.
For miscible polymer interfaces at times less than $t_c \sim 0.1$M$\tau$, there is a clean fracture along the initial interface.
The ends of chains that have diffused across the interface are simply pulled
out.
The small fracture energy is related to the friction resisting chain pull out.

For $t > t_c$ the interface is strong enough to support craze formation.
Direct examination of interfacial entanglements shows that the areal density $\Sigma$ of chains that have diffused far enough to form at least one topological constraint with chains on the opposing surface saturates at $t_c$.
This first TC is sufficient to transfer stress and stabilize the craze. 
However the interfacial region remains weak and for $t< t_s \sim 0.5$M$\tau$
the craze eventually fails through chain pullout near the interface.
For $t >t_s$ chains are entangled enough that craze breakdown produces chain scission.
This is the mechanism of craze failure in the bulk and 
the interfacial craze is as strong as the bulk for $t > t_b \sim 3$M$\tau$.
This is the same time where the interfacial shear strength saturates
and coincides with the recovery of a bulk density of entanglements in the
interfacial region \cite{ge13}.


The interfacial fracture energy $G_I$ is calculated by using our simulation
results as inputs to Brown's fracture model. As in experiment, $G_I$ increases roughly as $t^{1/2}$ before saturating at the bulk fracture energy $G_b$.
A linear correlation between $G_I$ and the density of interfacial entanglements is observed until $G_I$ saturates at $G_b$.
A similar linear relation between shear strength and interfacial entanglement density was observed previously \cite{ge13,ge13b}.
However the increase in shear strength is only a factor of 3 while the fracture energy increases by two orders of magnitude.

We also simulate tensile failure at immiscible polymer interfaces. Due to the lack of entanglements across the interface, the dominant failure mechanism is chain pullout at the interface without stable craze formation. This reduces the interfacial fracture energy $G_I$. However, once the interfacial width rises above the threshold value for entanglement development, $G_I$ is dramatically increased.

The above studies are for the simplest polymer model that captures the effects
of entanglements.
In a recent study of thermal healing of cracks we considered the effect
of increasing chain stiffness and thus lowering the entanglement length.
The results show that interchain friction may compete with entanglements
in determining interfacial strength \cite{ge14}.
We hope that our findings spur further studies with more realistic potentials
and can be used to improve existing macroscopic fracture models of polymer glasses and interfaces.


\begin{acknowledgement}
We thank E. Kramer, D. Perahia and M. Rubinstein for useful discussions and T. O'Connor for creating final versions of Figs. 2 and 6.
This work was supported by the National Science Foundation under 
grants DMR-1006805, DMR-1309892, CMMI-0923018, and OCI-0963185.
MOR acknowledges support from the Simons Foundation.
This research used resources at the National Energy Research Scientific Computing Center (NERSC), which is supported by the Office of Science of the United States Department of Energy under Contract No. DE-AC02-05CH11231. Research was carried out in part, at the Center for Integrated Nanotechnologies, a U.S. Department of Energy, Office of Basic Energy Sciences user facility. Sandia National Laboratories is a multi-program laboratory managed and operated by Sandia Corporation, a wholly owned subsidiary of Lockheed Martin Corporation, for the U.S. Department of Energy's National Nuclear Security Administration under contract DE-AC04-94AL85000.
\end{acknowledgement}

\bibliography{thesis}

\begin{mcitethebibliography}{52}
\bibitem[foo()]{foot1}
If $r_c$ was not reduced, the density and chain conformation changes would be
  much larger than in experiments because the melt is at a very elevated
  temperature to speed diffusion.\relax
\mciteBstWouldAddEndPunctfalse
\mciteSetBstMidEndSepPunct{\mcitedefaultmidpunct}
{}{\mcitedefaultseppunct}\relax
\EndOfBibitem
\end{mcitethebibliography}

\end{document}